\documentclass[twocolumn,showpacs,preprintnumbers,amsmath,amssymb]{revtex4}

\usepackage{graphicx}
\usepackage{dcolumn}
\usepackage{bm}

\begin{document}

\title{A general model for collaboration networks}
\author{Tao Zhou}
\author{Ying-Di Jin}
\author{Bing-Hong Wang}
\email{bhwang@ustc.edu.cn}
\affiliation{Nonlinear Science Center
and Department of Modern Physics, University of Science and
Technology of China, Hefei Anhui, 230026, PR China }
\author{Da-Ren He}
\author{Pei-Pei Zhang}
\author{Yue He}
\author{Bei-Bei Su}
\affiliation{College of Physics Science and Technology, Yangzhou
University, Yangzhou Jiangsu, 225002, PR China }
\author{Kan Chen}
\affiliation{ Department of Computational Science, Faculty of
Science, National University of Singapore, Singapore 117543}
\author{Zhong-Zhi Zhang}
\affiliation{Institute of System Engineering, Dalian University of
Technology, Dalian Liaoning, 116024 PR China}

\date{\today}

\begin{abstract}
In this paper, we propose a general model for collaboration
networks. Depending on a single free parameter ``{\bf preferential
exponent}", this model interpolates between networks with a
scale-free and an exponential degree distribution. The degree
distribution in the present networks can be roughly classified
into four patterns, all of which are observed in empirical data.
And this model exhibits small-world effect, which means the
corresponding networks are of very short average distance and
highly large clustering coefficient. More interesting, we find a
peak distribution of act-size from empirical data which has not
been emphasized before of some collaboration networks. Our model
can produce the peak act-size distribution naturally that agrees
with the empirical data well.
\end{abstract}


\maketitle

\section{Introduction}
The last few years have witnessed a tremendous activity devoted to
the characterization and understanding of complex
networks\cite{Reviews1,Reviews2,Reviews3,Reviews4}, which arise in
a vast number of natural and artificial systems, such as
Internet\cite{Internet1,Internet2,Internet3}, the World Wide
Web\cite{WWW1,WWW2}, social networks of acquaintance or other
relations between individuals\cite{Social1,Social2,Social3},
metabolic networks\cite{Metabolic1,Metabolic2,Metabolic3}, food
webs\cite{Foodwebs1,Foodwebs2,Foodwebs3,Foodwebs4} and many
others\cite{Others1,Others2,Others3,Others4,Others5,Others6,Others7}.
Owing to the computerization of data acquisition process and the
availability of high computing powers, scientists have found that
the networks in various fields have some common characteristics,
which inspires them to construct a general model. Recently, some
pioneer works have been done that bring us new eyes of the
networks' evolution mechanism. For instance, Barab\'{a}si and
Albert's introduced a scale-free network model (BA
network)\cite{BA}, which suggests that two main ingredients of
self-organization of a network in a scale-free structure are
growth and preferential attachment.

So far, BA model may be the most successful model to fit the
empirical results of complex system, but there are still a great
number of real networks whose evolution mechanisms cannot be
explained by BA model. For truth, we should not ask for an
all-powerful model which can explain the reason of a freewill real
network coming into being, since many different networks have
distinct underlying growth mechanism. Therefore, it is meaningful
to construct a microscopic suitable model aiming at a special kind
of networks.

A particular class of networks is the so-called collaboration
networks, which is considered to be a kind of social networks in
the early studies. In the social science literatures, a
collaboration network is generally defined as a network of actors
connected by common membership in group of some sort, such as
clubs, teams or organizations. Some empirical studies relevant to
collaboration networks have been done, including scientific
collaboration
networks\cite{Newman2001PRE,Newman2001PNAS,Fan2004,Li2004,MSN},
board of directorships\cite{Directorships1,Directorships2}, movie
actors collaboration networks\cite{WS1998}, social events
attending networks for women\cite{Women}, and so on. It is
worthwhile to point out that the extension of collaboration
networks should not be restricted within social networks, one
instance is the software collaboration networks\cite{Myers2003},
and we will show more examples of collaboration networks irrelated
to social networks in the following text.

Ramasco, Dorogovtsev and Pastor-Satorras have proposed a model for
collaboration networks(RDP model)\cite{RDP}. In RDP model, They
found the power law behavior in degree distribution, the
nontrivial clustering-degree correlation and nontrivial
degree-degree correlation. Very recently, Li et al have
established a model for weighted collaboration networks in which
both the power law weight distribution and degree distribution are
obtained\cite{Li2005}.

In this paper, we propose a general model for collaboration
networks. Depending on a single free parameter, the preferential
exponent, this model interpolates between networks with a
scale-free and an exponential degree distribution.  The degree
distributions of the present networks can be roughly classified
into four patterns, and all of them are observed in empirical
data. And this model exhibits small-world effect, which means the
corresponding networks are of very short average distance and
highly large clustering coefficient. More interesting, we find a
peak distribution of act-size from empirical data which has not
been emphasized before of some collaboration networks. Our model
can produce the peak act-size distribution naturally that agrees
with the empirical data well.

The present paper is organized as follows. In section 2, we
introduce a simple and general model for collaboration networks.
In section 3, we show the small-world effect exhibited by this
model. In section 4, we display the simulation results of degree
distribution, and demonstrate that the degree distribution
approximate to stretched exponential distribution\cite{SED} with
adjustable parameter $c$. In section 5, we show the simulation
results and some empirical data of act-size distribution. The
comparison and qualitative discussion are also included. Finally,
in section 6, we draw the main conclusion of our work.

\section{The model}
Our network starts with $m_0$ nodes which are fully connected.
Then, at each time step, we add a new node into the network which
will have a collaboration with some existing nodes. Inspired by BA
model, we assume that the probability that an existing node $x$ is
chosen to be an actor in the collaboration is proportional to
$k^{\alpha} (\alpha \geq 0)$, where $k$ is the degree of $x$ and
$\alpha$ is the so-called ``preferential exponent" denoting the
degree of preferential attachment. For $\alpha >0$, we have
preferential attachment.

All the existing nodes which are chosen to be collaborated will
link to the new node. That is to say if two chosen old nodes have
never collaborated so far, there will be a new edge added
connecting them. It is obvious that this model can be stretched to
weighted one by using the times of collaborations between the
corresponding two nodes as edge weight. Since the aim of this
paper is to introduce the characteristics of non-weighted
networks, the simulation and analysis relevant to weighted
networks will not be included, which will be published elsewhere.

It should be taken note to that the act-size is not fixed at each
time step since whether a certain node is chosen will not affect
other nodes. We suppose that a node with degree
 $k$ will be chosen with the probability
\begin{equation}
\pi (k)=\lambda k^{\alpha}/\sum_{i}k_{i}^{\alpha}
\end{equation}
in which $\lambda$ is a constant, and $\sum _{i}k_{i}^{\alpha}$ is
the normalization factor. Using $<s>$ presenting the average value
of act-size such as the mean number of authors per paper, we will
conclude that $<s>\approx \lambda+1$, since the number of nodes
chosen each time has the expecting value $\Sigma \pi (k)=\lambda$.
Thus, $\lambda$ is a parameter which can be used to control the
average act-size of the whole network. Therefore, when we simulate
an idiographic network of known average act-size, the parameter
$\lambda$ is fixed.

\section{Small World Effect}
In a network, the distance between two nodes is defined as the
number of edges along the shortest path connecting them. The
average distance $L$ of the network, then, is defined as the mean
distance between two nodes, averaged over all pairs of nodes and
often considered to be one of the most important parameters to
measure the efficiency of communication networks. The clustering
coefficient $C(x)$ of node $x$ is the ratio between the number of
edges among $A(x)$ and the total possible number, where $A(x)$
denotes the set of all the neighbors of $x$. The clustering
coefficient $C$ of the whole network is the average of $C(x)$ over
all $x$. Empirical studies indicate that most real-life networks
have much smaller average distance (as $L\sim\ln N$ where $N$ is
the number of nodes in the network) than the completely regular
networks and much greater clustering coefficient than those of the
completely random networks. And these two properties, small
average distance and large clustering coefficient, make up of the
so-called small world effect.

\begin{figure}
\scalebox{0.8}[0.9]{\includegraphics{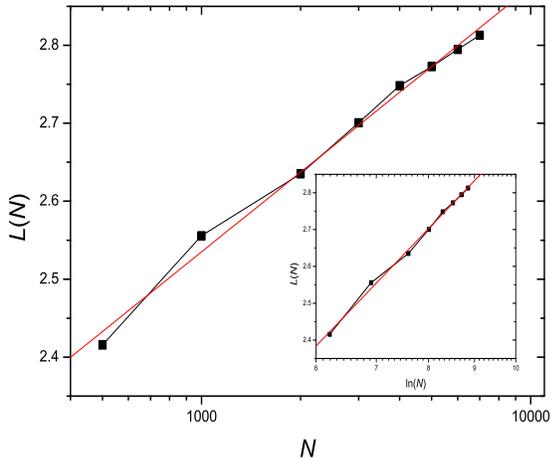}}
\caption{\label{fig:epsart} The dependence between the average
distance $L$ and the network size $N$. One can see that $L$
increases very slowly as $N$ increases. The main plot exhibits the
curve where $L$ is considered as a function of $\texttt{ln}N$,
which is well fitted by a straight line. The curve is above the
fitting line when $N\leq 4000$ and under the fitting line when
$N\geq 5000$, which indicates that the increasing tendency of $L$
can be approximated as $\texttt{ln}N$ and in fact a little slower
than $\texttt{ln}N$. The inset shows the average distance $L$ vs
$\texttt{lnln}N$, the error of linear-fitting by form
$\texttt{lnln}N$ is smaller than $\texttt{ln}N$, indicating that
the networks may be considered as ultrasmall world
networks\cite{Ultrasmall}. All the data are obtained by 10
independent simulations with parameters $\alpha=1$ and
$\lambda=3$.}
\end{figure}

Inspired by the empirical studies on real-life networks, Watts and
Strogatz proposed a one-parameter model(WS model) that
interpolates between an ordered finite-dimensional lattice and a
random graph by randomly rewiring each edge of the regular lattice
with probability $p$\cite{WS1998}. In WS model, $L$ scales
logarithmatically with $N$, and the clustering coefficient
decrease with $N$, which is in excellent agreement with the
characteristics of real networks. The pioneering article of Watts
and Strogatz started an avalanche of research on the properties of
small-world networks. In this section, we would like to
demonstrate that the networks generated by the present rules
display small-world effect.

At first, we study the average distance of the present model using
the approach similar to that in references\cite{RAN1,RAN2}. Using
symbol $d(i,j)$ to represent the distance between nodes $i$ and
$j$, the average distance of present networks with order $N$,
denoted by $L(N)$, is defined as:
\begin{equation}
L(N)=\frac{2\sigma (N)}{N(N-1)}
\end{equation}
where the total distance is:
\begin{equation}
\sigma (N)=\sum_{1\leq i<j\leq N}d(i,j)
\end{equation}
The distance between two existing nodes will not increase with the
increasing of network size $N$, thus we have:
\begin{equation}
\sigma(N+1)\leq \sigma(N)+\sum_{i=1}^N d(i,N+1)
\end{equation}
Assume that $h$ existing nodes, $x_1, x_2, \cdots, x_h$, are
selected to collaborate with the new node $N+1$, then $d(i,N+1)$
is equal to the minimal distance between $i$ and any one of the
$h$ nodes plus 1:
\begin{equation}
d(i,N+1)=\texttt{min}\{d(i,x_j)|j=1, 2, \cdots ,h\}+1
\end{equation}
In a rough version, the sum
$\sum_{i=1}^N=\texttt{min}\{d(i,x_j)\}$ can be expressed
approximately in terms of $L(N-h+1)$:
\begin{equation}
\sum_{i=1}^N=\texttt{min}\{d(i,x_j)\}\approx(N-h)L(N-h+1)
\end{equation}
In order to avoid the network being unconnected, we always set
$\lambda>1$ and compel $h\geq 1$, which leads to
$(N-h)L(N-h+1)\leq(N-1)L(N)$.

Combining those results above, we have:
\begin{equation}
\sigma(N+1)<\sigma(N)+N+\frac{2\sigma(N)}{N}
\end{equation}
Consider (7) as an equation, then the increasing tendency of
$\sigma(N)$ is determined by the equation:
\begin{equation}
\frac{d\sigma(N)}{dN}=N+\frac{2\sigma(N)}{N},
\end{equation}
which leads to
\begin{equation}
\sigma(N)=N^2\texttt{ln}N+H
\end{equation}
where $H$ is a constant. As $\sigma(N)\sim N^2L(N)$, we have
$L(N)\sim \texttt{ln}N$. Which should be pay attention to, since
(7) is an inequality in fact, the precise increasing tendency of
$L$ may be a little tardier than $\texttt{ln}N$.

In figure 1, we report the typical simulation result on average
distance of the present networks under parameters $\alpha=1$ and
$\lambda=3$, which agrees with the analytic result well.

\begin{figure}
\scalebox{0.8}[0.9]{\includegraphics{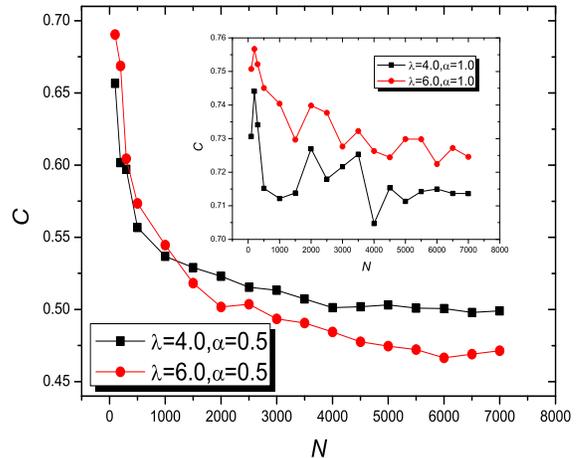}}
\caption{\label{fig:epsart} The clustering coefficient vs network
size $N$. The main plot and inset exhibit the dependence between
the clustering coefficient $C$ and network size $N$. One can see
clearly that the clustering coefficient of the present networks is
sufficient large even for big $N$.}
\end{figure}

\begin{figure}
\scalebox{0.8}[0.9]{\includegraphics{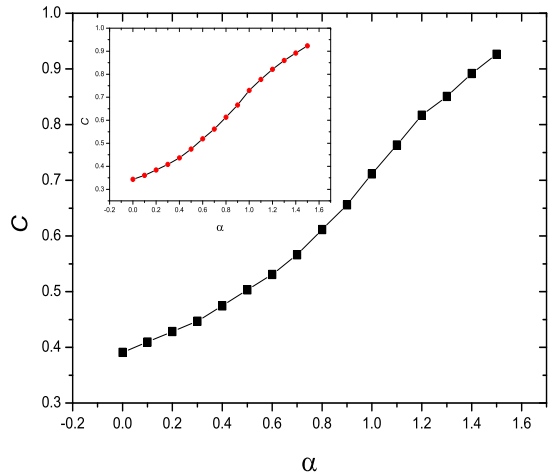}}
\caption{\label{fig:epsart} The clustering coefficient vs
preferential exponent $\alpha$. The two curves can be considered
as the clustering coefficient as a function of $\alpha$ with
network size $N=5000$ fixed, which increase monotonically with the
increasing of $\alpha$. The main plot and inset are of
$\lambda=4.0$ and $\lambda=6.0$ respectively. It is clear that the
clustering coefficient is sensitive to the preferential exponent
but influenced little by $\lambda$.}
\end{figure}

In succession, let's discuss the clustering coefficient. As we
mentioned above, for an arbitrary node $x$, the clustering
coefficient $C(x)$ is:
\begin{equation}
C(x)=\frac{2E(x)}{k(x)(k(x)-1)}
\end{equation}
where $E(x)$ is the number of edges between any two nodes in the
neighbor-set $A(x)$ of node $x$, and $k(x)=|A(x)|$ denotes the
degree of node $x$. The clustering coefficient $C$ of the whole
network is defined as the average of $C(x)$ over all nodes.

In figure 2, we report the simulation results on clustering
coefficient of the present networks vs network size. From figure
2, one can find that the clustering coefficient of the present
networks is sufficient large even for big $N$. Therefore, our
model exhibits completely different clustering structure from that
of BA networks, in which the clustering coefficient is very small
and decreases with the increasing of network size $N$, following
approximately $C\sim \frac{\texttt{ln}^2N}{N}$\cite{Klemm}.

In addition, we plot the clustering coefficient as a function of
$\alpha$ with network size $N=5000$ fixed in figure 3. The two
curves with different $\lambda$ are almost the same, thus the
clustering coefficient is influenced little by $\lambda$. Both of
the two curves increase monotonically with $\alpha$, since
$\alpha$ represents the degree of preferential attachment, this
phenomenon reveals that the huger difference between attraction of
preponderant and puny individuals will lead to greater clustering
behavior.

Even when the network grows without preferential attachment(i. e.
$\alpha=0$), the clustering coefficient of our model is much
greater than completely random networks, because of its special
linking mode proposed here. For $\alpha>1.5$, the clustering
coefficient approximate 1, and the structure of corresponding
networks is similar to a star in topology\cite{GT1,GT2}. The
difference is that in our networks with very large $\alpha$, the
central part are not one node like star, but many nodes almost
fully connected to each other. Since the structure for networks
with $\alpha>1.5$ is much different from reality, we will not
discuss their characteristics hereinafter.

Summing up, the present networks own both very large clustering
coefficient and very small average distance which agree with
previous empirical studies well.

\section{Degree distribution}
The degree distributions of real-life networks are
various\cite{Ameral2000,Strogatz2001}. Some of them such as
acquaintance network of Mormons\cite{Bernard1988} are Guassian;
some such as power-grid of southern California are
exponential\cite{WS1998} and some such as network of World Wide
Web are power-law\cite{WWW1,WWW2}. However, the degree
distributions of most real-life networks do not obey these simple
forms above, they may interpolate between Guassian and exponential
ones such as the network of world airports\cite{Ameral2000}, or
interpolate between exponential and power-law ones such as
citation networks in high energy physics\cite{Lehmann2003}, or in
another form\cite{Scala2000}.

\begin{figure}
\scalebox{0.8}[0.9]{\includegraphics{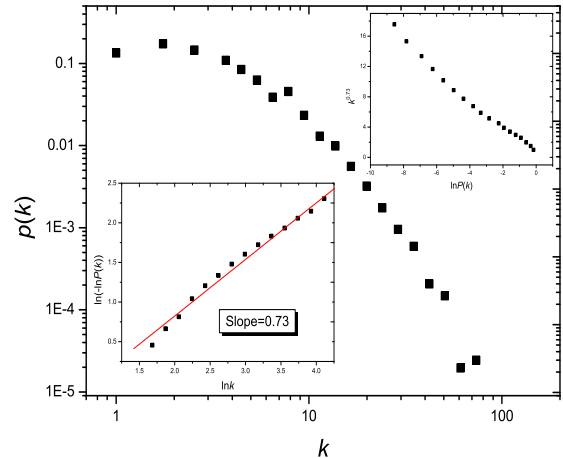}}
\caption{\label{fig:epsart} The degree distribution of scientific
collaboration network, where $p(k)$ denotes the probability a
randomly selected node is of degree $k$ and $P(k)=\int_k^\infty
p(k)dk$ is the cumulation probability. The main plot is the degree
distribution. The left-down inset shows how the quantity
$\texttt{ln}(-\texttt{ln}P(k))$ behaviors as a function of
$\texttt{ln}k$, which can be approximately fitted by a straight
line of slope $0.73\pm 0.02$, thus these data obey SED of
$c\approx 0.73$(see Equ.A3). The right-up inset exhibits $k^c$ vs
$\texttt{ln}P(k)$, which approximates to a line with negative
slope.}
\end{figure}

\begin{figure}
\scalebox{0.8}[0.9]{\includegraphics{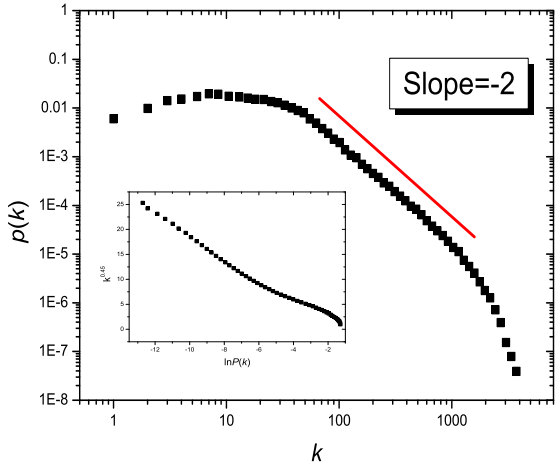}}
\caption{\label{fig:epsart} The degree distribution of actors
collaboration network. The main plot is the degree distribution
that displays power-law only in the interval about $k\in
[50,1000]$. The solid line is of slope -2 for comparison. These
data can be approximately fitted by SED of $c=0.45\pm 0.02$. The
inset exhibits $k^c$ vs $\texttt{ln}P(k)$.}
\end{figure}

In this section, we focus on the empirical results about
collaboration networks. About four years ago, Newman investigated
the statistic properties of scientific collaboration networks. He
demonstrated that the degree distribution can be well fitted by an
truncated power-law in the form $p(k)\sim
k^{-\tau}e^{-\frac{k}{k_c}}$\cite{Newman2001PNAS}, or considered
as a double power-law\cite{Newman2001PRE}. Cs\'{a}nyi and
Szendr\H{o}i have investigated the acquaintance networks from WIW
project where the double power-law is also detected\cite{MSN}. In
fact, Lehmann et al have shown an example where the observed
double power-law can be well fitted by a stretched exponential
form\cite{Lehmann2003}. In Appendix A, the details about stretched
exponential distribution(SED) is shown, including the definition
of SED, the basic properties of SED, the relations between SED and
exponential distribution as well as power-law distribution, and
the reason why we use SED in this paper. Figure 4 shows the degree
distribution of scientific collaboration network proposed by
Newman\cite{Newman2001PRE,Newman2001PNAS} which can be well fitted
by SED with $c=0.73$ indicating this distribution is more
approximated to exponential form rather than power-law form.
Another famous example is the collaboration network of movie
actors\cite{WS1998}, which displays power-low only in its middle
region. This distribution ia also consistent with a stretched
exponential form with $c=0.45$(see figure 5).

We did some empirical work on collaboration networks as well as
theoretical work and found that the degree distributions of many
real-life collaboration networks in various fields approximately
obey the stretched exponential form\cite{Zhang2005}. For example,
if we consider the traveling sites as actors and the traveling
routes that contain several sites as acts, then the recommended
traveling routes from the web {\it Walkchina} and {\it Chinavista}
in the year 2003 will form a Chinese touristry collaboration
network, whose degree distribution is accurately consistent with
SED of $c=0.50$\cite{Zhang2005,Zhang2004}.

\begin{figure}
\scalebox{0.8}[0.9]{\includegraphics{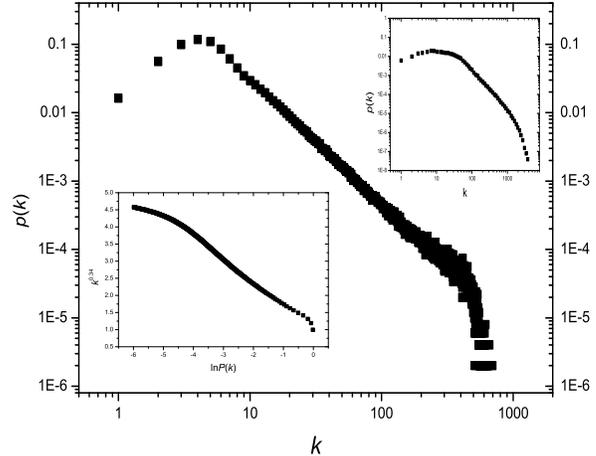}}
\caption{\label{fig:epsart} A typical simulation result of degree
distribution with $N=5000$, $\alpha=1.0$ and $\lambda=4$. The main
plot is the average of 100 independent simulations. The degree
distribution exhibits observed power-law behavior in its middle
region, which is similar to the case of movie actors collaboration
network(see figure 5 or the right-up inset for comparison). The
left-down inset shows $k^{0.34}$ vs $lnP(k)$, which is
approximated to a negative line indicating that the corresponding
degree distribution can be well fitted by stretched exponential
form of $c=0.34$.}
\end{figure}

\begin{figure}
\scalebox{0.8}[0.9]{\includegraphics{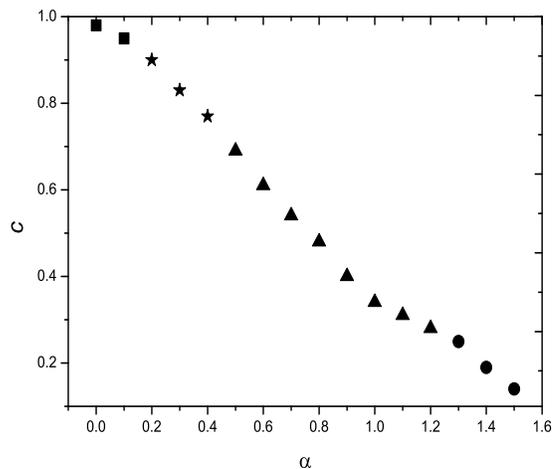}}
\caption{\label{fig:epsart} The parameter $c$ of SED as a function
of preferential exponent $\alpha$. The value of $c$ monotonously
decreases from 0.98 to 0.14 as the increasing of $\alpha$. All the
data are the average of 100 independent simulations, where
$N=5000$ and $\lambda=4$ are fixed. The pattern of degree
distributions for different $\alpha$ can be roughly classified
into four types: exponential($\blacksquare$),
arsy-varsy($\bigstar$), semi-power law($\blacktriangle$) and power
law($\bullet$).}
\end{figure}

\begin{figure}
\scalebox{1}[1]{\includegraphics{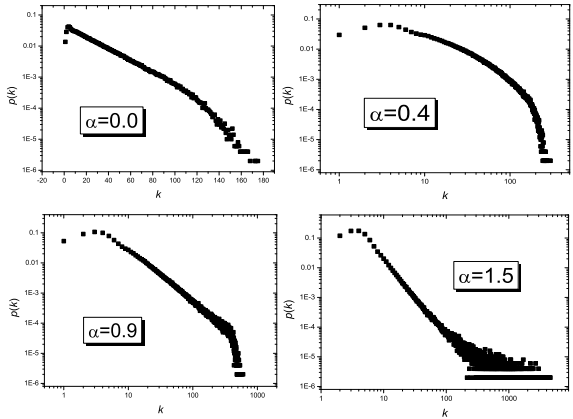}}
\caption{\label{fig:epsart} The representative instances for the
four patterns of degree distribution. {\bf
Exponential}($\alpha=0.0$), the degree distribution obeys
exponential form except its tail; {\bf arsy-varsy}($\alpha=0.4$),
the degree distribution does not exhibit observed exponential or
power law; {\bf semi-power law}($\alpha=0.9$), the degree
distribution exhibits observed power-law behavior only in its
middle region; {\bf power law}($\alpha=1.5$), the degree
distribution displays power law in all the region except a ridgy
head and a fat tail.}
\end{figure}

In succession, let's discuss the degree distribution of the
networks generated from our model. Since the number of both the
selected nodes and new edges are unfixed during each time step, it
is hard for us to obtain the analytic results. For comparison, we
will give analytic results for a special case of this model in
Appendix B, and here, only the numerical results are shown. In
figure 6, we report a typical simulation result with $\alpha=1.0$
and $\lambda=4$. The degree distribution is similar to that of
movie actors collaboration network and can be well fitted by
stretched exponential form of $c=0.34$. We have also investigated
how the two parameters affect the degree distribution. In figure
7, one can see that the parameter $c$ of SED monotonously
decreases from 0.98 to 0.14 with $\alpha$, the smaller $\alpha$
corresponds ``more exponential" network while the larger one
corresponds ``more power-law" one. As we mentioned above, for
$\alpha>1.5$, the networks are star-like in which the degree of
hub node(i.e. the node of maximal degree) will exceed half of the
network size, which has not been observed in the real-life
collaboration networks, and will not be discussed hereinafter
wither. To have an intuitionistic sight into the degree
distribution of the present networks, we roughly classify those
distributions into four patterns. They are {\bf Exponential}, {\bf
arsy-varsy}, {\bf semi-power law} and {\bf power law}. In figure
8, we show the representative instances for the four patterns.
There are no unambiguous borderline between two neighboring
patterns. We also have checked that the parameter $\lambda$ affect
the holistic property of degree distribution little; the larger
$\lambda$ only makes the head larger for sufficient big $N$.

In a word, many real-life collaboration networks are of degree
distribution lying between exponential and power-law ones that can
be well fitted by stretched exponential form, and the present
model can generate networks of degree distribution from ``almost
exponential" to ``almost power-law" containing four patterns.

\section{Act-size distribution}
Act-size distribution is another characteristic distribution
beside degree distribution for collaboration networks, which is a
particular distribution of collaboration networks. In many
real-life cases, this distribution is single-peaked, and decays
exponentially. One famous instance is the networks of corporate
directors\cite{Davis1996} in which the act-size distribution,
defined as the number of directors per board, is single-peaked(see
figure 8 in ref. \cite{Strogatz2001}). We have also done some
empirical works about act-size distribution of collaboration
networks\cite{Zhang2005}. All of these networks, including Chinese
touristry collaboration network, bus route network, scientific
collaboration network, and so on, exhibit single-peaked act-size
distribution. In figure 9, we show two examples, Chinese touristry
collaboration network and scientific collaboration network. In the
former case, the act-size is the number of traveling sites per
traveling route; the latter one only contains the 2062 papers in
Vol. 93 of {\it Physical Review Letters}, where each paper is
considered as an act and the act-size is the number of authors.
There are 98 papers having authors more than 20, which have not
been shown in figure 9. Both of the two distributions are
single-peaked and in an approximately exponential form.

\begin{figure}
\scalebox{0.41}[0.5]{\includegraphics{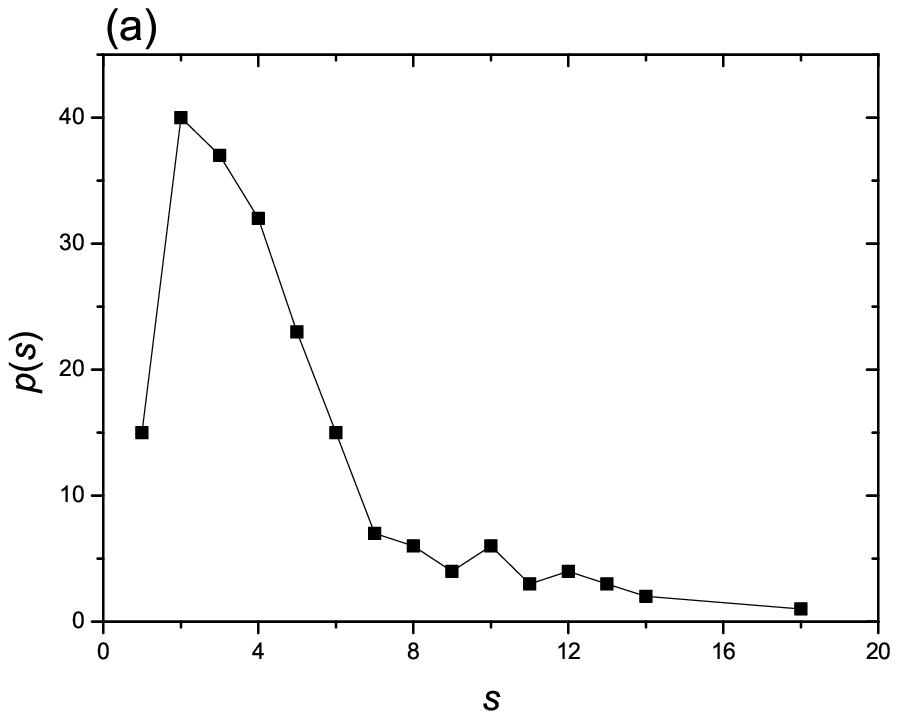}}
\scalebox{0.41}[0.5]{\includegraphics{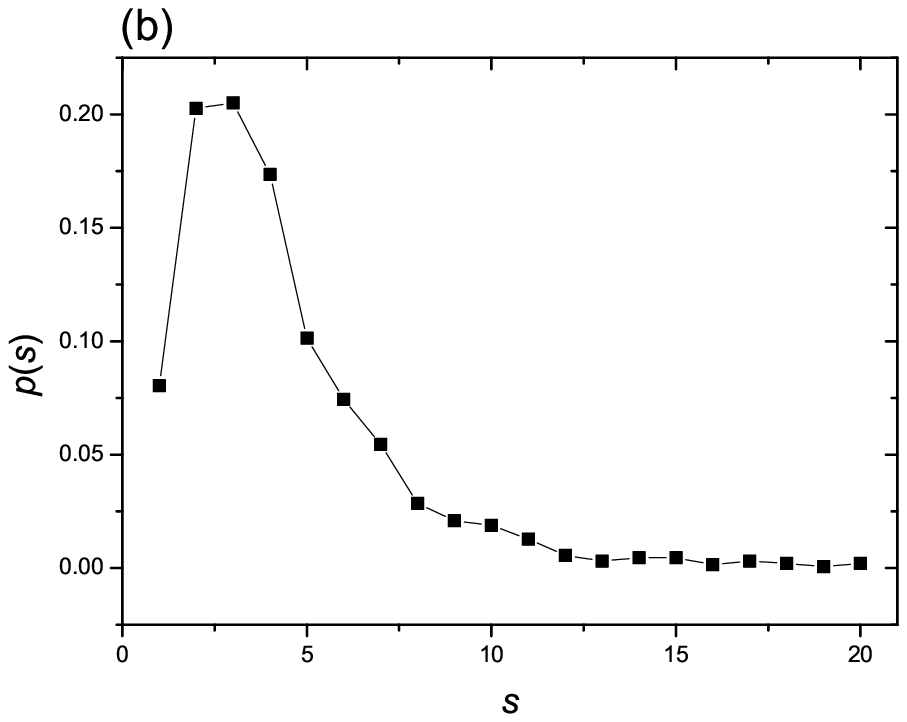}}
\scalebox{0.41}[0.5]{\includegraphics{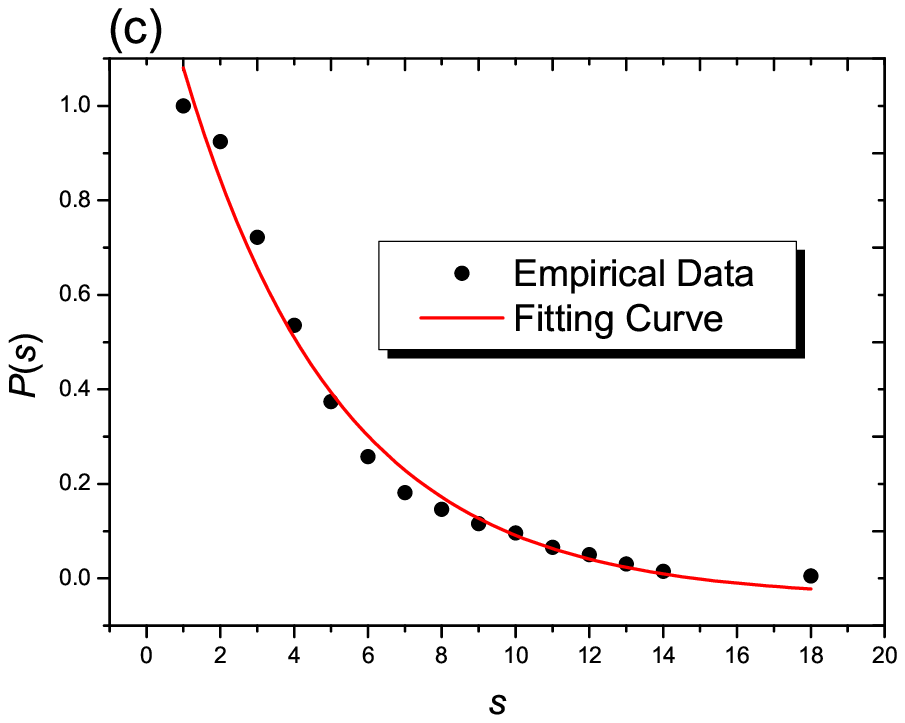}}
\scalebox{0.41}[0.5]{\includegraphics{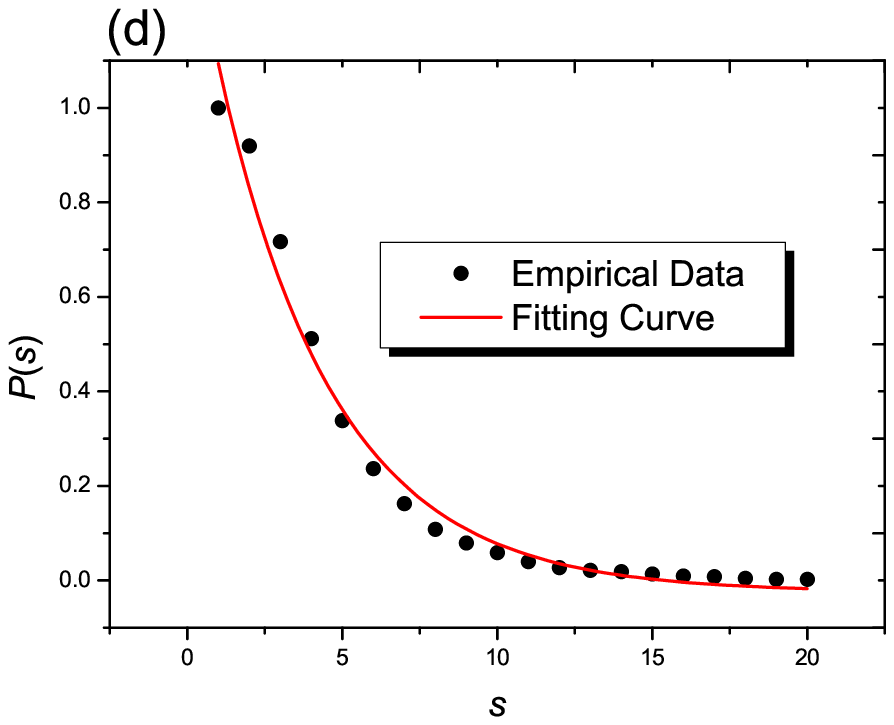}}
\caption{\label{fig:epsart} Empirical results about act-size
distribution. Figure {\bf a}\&{\bf b} show the act-size
distribution of Chinese touristry collaboration network and the
scientific collaboration network of {\it Physical Review Letters},
respectively. Both the two distributions display obviously
single-peaked behavior. Figure {\bf c}\&{\bf d} are the
corresponding cumulation distributions for those two networks. The
red solid curves are the fitting curves of exponential form. In
these four plots, the symbol $s$, $p(s)$ and $P(s)=\int_s^\infty
p(s)ds$ denote act-size, the probability that a randomly selected
act are of size $s$, and the cumulation probability,
respectively.}
\end{figure}

However, the act-size distribution seems not as attractive as
degree distribution, thus the observed peaked behavior has not
been emphasized before. It is always ignored\cite{Li2005}, or only
considered as an extrinsical factor\cite{RDP}, having nothing to
do with and not being affected by the evolutionary mechanism of
networks. In our model, the act-size distribution is not generated
based on a static perspective like the degree distribution of
configuration model\cite{Configuration}, but an indiscerptible
part of the dynamical mechanism of network evolution. It is clear
that, when $\alpha=0$, for sufficient large $N$, the act-size
distribution is Possionian distribution, single-peaked and
decaying approximately exponentially. Contrary to the case of
degree distribution, the numerical study indicates that the
act-size distribution is insensitive to $\alpha$.

\begin{figure}
\scalebox{0.8}[0.9]{\includegraphics{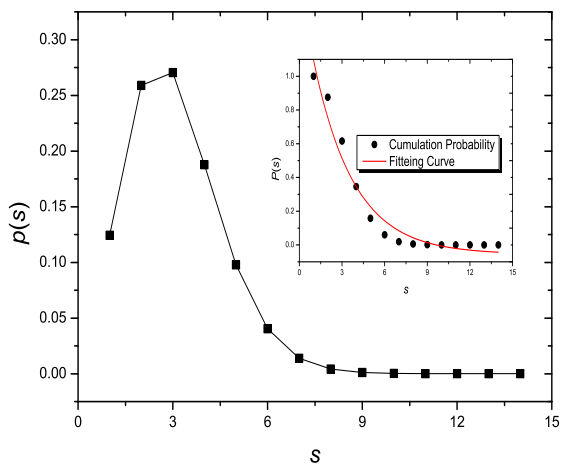}}
\caption{\label{fig:epsart} A typical simulation result on
act-size distribution with $N=5000$, $\alpha=0.4$ and
$\lambda=2.0$. The data are the average of 100 independent
simulations. The main plot exhibits obviously single-peaked
behavior. The inset shows the corresponding cumulation
distribution, which can be well fitted by an exponential
function(see the red solid curve).}
\end{figure}

A typical simulation result is shown in figure 10, one can compare
this to the empirical data for scientific collaboration
networks(see figure {\bf 9b}\&{\bf 9d}). We set $\alpha=0.4$ since
it makes the parameter $c$ of the two networks pretty much the
same thing. Clearly, the act-size distribution generated by our
model is well consistent to the real-life one qualitatively.

\section{Conclusion}
In summary, we have constructed a general model for collaboration
networks, the basic constituents of which are preferential
attachment and particular selecting and linking rules aiming at
collaboration networks. The present networks are both of very
large clustering coefficient and very small average distance,
which is consistent with the previous empirical results that
collaboration networks display small-world effect. We argue that,
the degree distribution of many real-life collaboration networks
may appropriately be fitted by stretched form. Numerical study
indicates the degree distribution of the present networks can be
well fitted by stretched form with the parameter $c$ decreaing
from 0.98 to 0.14 as the increasing of $\alpha$. We roughly
classify the degree distribution of our model into four patterns,
{\bf Exponential}(bus route network in Beijing\cite{Zhang2005}),
{\bf arsy-varsy}(scientific collaboration
network\cite{Newman2001PRE,Newman2001PNAS}), {\bf semi-power
law}(movie actors collaboration network\cite{WS1998} and Chinese
touristry collaboration network\cite{Zhang2005,Zhang2004}) and
{\bf power law}(bus route network in Yangzhou\cite{Zhang2005}),
all of which are observed in the empirical data. More absorbing,
we find the act-size distribution is single-peaked and decaying
exponentially, which can be reproduced by our model naturally.

Although this model is too simple and rough, it offers a good
starting point to explain the existing empirical data and can be
easily extended when more factors that may affect network
evolution are considered. In addition, it is obvious that this
model can be stretched to weighted network model if the edge
weight is used to represent the times of collaborations between
the corresponding two nodes. The further statistical properties of
the present networks, such as the degree-degree correlation, the
clustering-degree correlation and so on have also been
investigated, which will be published elsewhere.

\begin{acknowledgments}
We thank Prof. Zeng-Ru Di for his irradiative talk. BHWang
acknowledges the support by the National Natural Science
Foundation of China(NNSFC) under No. 70271070, 10472116 and the
Specialized Research Fund for the Doctoral Program of Higher
Education under No. 20020358009. DRHe acknowledges the support by
NNSFC under No. 70371071. KChen acknowledges the support by the
National University of Singapore research grant R-151-000-028-112.
TZhou acknowledges the support by NNSFC under No. 70471033 and the
Foundation for Graduate Students of University of Science and
Technology of China under Grant No. KD200408.
\end{acknowledgments}

\appendix

\section{Power law and Stretched Exponential Distribution}

Frequency or probability distribution functions(PDF) that decay as
a power law have acquired a special status in the last decade. A
power law distribution $p(x)$ characterizes the absence of a
characteristic size: in dependently of the value of events $x$. In
contrast, an exponential for instance or any other functional
dependence does not enjoy this self-similarity. In words, a power
law PDF is such that there is the same proportion of smaller and
larger events, whatever the size one is looking at within the
power law range. Since the power law distribution has repeatedly
been claimed to describe many natural phenomena and been proposed
to apply to a vast set of social an economic
statistics\cite{Manderbrot1983,Mantegna1995,Xie2004,Bak1994,Wang2001,Newman2004},
it is considered as one of the most striking signatures of complex
dependence. Empirically, a power law PDF is represented by a
linear dependence in a in a double logarithmic axis plot of the
frequency or cumulative number as a function of size. However,
logarithms are notorious for contraction data and the
qualification of a power law is not as straight-forwards as often
believed. In addition, log-log plots of data from natural
phenomena in nature and economy often exhibit a limit linear
regime followed by a signature curvature. Latherr\`{e}re and
Sornette\cite{SED} explore and test the hypothesis that the
curvature observed in log-log plots of distribution of several
data sets taken from natural and economic phenomena might result
from a deeper departure from the power law paradigm and might call
for an alternative description over the whole range of the
distribution. Thus, they propose a {\bf stretched exponential
distribution}(SED):
\begin{equation}
p(x)dx=c(x^{c-1}/x_{0}^{c})\texttt{exp}[-(x/x_{0})^{c}]dx
\end{equation}
such that the cumulation distribution is
\begin{equation}
P(x)=\texttt{exp}[-(x/x_{0})^{c}]
\end{equation}

Stretched exponentials are characterized by an exponent $c$
smaller that one. The borderline $c=1$ corresponds to the usual
exponential distribution. For $c$ smaller than one, the
distribution presents a clear curvature in a log-log plot. Based
on the reasons discussed above, we use the SED to fit the degree
distribution of our model. In the simulations, using the frequency
$p(k)$ of degree $k$ as $x$-axis and $k^{c}$ as $y$-axis, we will
obtain a line with negative slope, if the degree satisfy strict
Stretched Exponential Distribution.

In numerical case, write down the equivalent form of Equ.(A2):
\begin{equation}
\texttt{ln}(-\texttt{ln}P(k))=c\texttt{ln}k-c\texttt{ln}k_0
\end{equation}
Using $\texttt{ln}k$ as $x$-axis and
$\texttt{ln}(-\texttt{ln}P(k))$ as $y$-axis, if the corresponding
curve can be well fitted by a straight line, then the slope will
be the value of $c$.

\section{A special model for collaboration networks of fixed act-size}
Under a very special case, the act-size is fixed. For example, if
the four players in a bridge game are considered as actors in one
act, then the act-size is 4. For comparison, in this appendix, we
introduce a resovable model for this special case.

This model starts with a $m$-complete network\cite{ex}, where
$m\geq 2$. At each time step, a new node is added and linked to
all the nodes of a randomly selected $m$-complete network. Under
these rules, not only the act-size, but also the number of new
edges in each time step is fixed, which makes the model very easy
to be analyzed.

Since after a new node is added to the network, the number of
$K_m$ increases by $m$, thus when the network is of order $N$, the
number of $K_m$ is $N_m=Nm-m^2+m$. Note that, when a given node's
degree increases one, the number of $K_m$ containing this node
increases $m-1$, hence for any node with degree $k$, it belongs to
$\phi_k=km-k-m^2$ $m$-complete networks. Let $n(N,k)$ be the
number of nodes with degree $k$ when $N$ nodes are present, now we
add a new node to the network, $n(N,k)$ evolves according to the
following rate equation\cite{RateEquation}:
\begin{equation}
n(N+1,k+1)=n(N,k)\frac{\phi_k}{N_m}+n(N,k+1)(1-\frac{\phi_{k+1}}{N_m})
\end{equation}
When $N$ is large enough, $n(N,k)$ can be approximated to $Np(k)$,
where $p(k)$ is the probability density function for the degree
distribution. In terms of $p(k)$, the above equation can be
rewritten as:
\begin{equation}
p(k+1)=\frac{N}{N_m}[p(k)\phi_k-p(k+1)\phi_{k+1}]
\end{equation}
Using the expression $p(k+1)-p(k)=\frac{dp}{dk}$, we can get the
continuous form of Eq.(B2):
\begin{equation}
p(k+1)+\frac{N}{N_m}[(km-k-m^2)\frac{dp}{dk}+(m-1)p(k+1)]=0
\end{equation}
Under the case $N\geq k\geq m$, this equation leads to
$p(k)\propto k^{-\gamma}$ with $\gamma=\frac{2m-1}{m-1}\in (2,3]$.
The simulation result accurately agrees with the analytic one for
large network size $N$.

In addition, there exists a bijection from node's degree to
clustering coefficient as:
\begin{equation}
C(k)=\frac{(m-1)(2k-m)}{k(k-1)}
\end{equation}
The clustering coefficient of the whole network can be obtained as
the mean value of $C(k)$ with respect to the degree distribution
$p(k)$:
\begin{equation}
C=\int_{k_{\texttt{min}}}^{k_{\texttt{max}}}C(k)p(k)dk,
\end{equation}
where $k_{\texttt{min}}=m$ is the minimal degree and
$k_{\texttt{max}}\gg k_{\texttt{min}}$ is the maximal degree.
Combine Eq.(B4) and Eq.(B5), note that
$p(k)=Ak^{\frac{2m-1}{m-1}}$ and
$\int_{k_{\texttt{min}}}^{k_{\texttt{max}}}Ap(k)dk=1$, we can get
the analytical result of $C$ by approximately treating
$k_{\texttt{max}}$ as $+\infty$. For example, when $m=2,3,4,5$ the
clustering coefficients are 0.739, 0.813, 0.851 and 0.875. Further
more, many real-life networks are characterized by the existence
of hierarchical structure\cite{Ravasz2003}, which can usually be
detected by the negative correlation between the clustering
coefficient and the degree. The BA network, which does not possess
hierarchical structure, is known to have the clustering
coefficient $C(x)$ of node $x$ independent of its degree $k(x)$,
while the present network has been shown to have $C(k)\sim
k^{-1}$, in accord with the observations of many real
networks\cite{Ravasz2003}.


\begin{thebibliography}{Reviews}
\bibitem{Reviews1} R. Albert and A. -L. Barab\'{a}si, Rev. Mod. Phys. {\bf 74}, 47(2002).
\bibitem{Reviews2} S. N. Dorogovtsev and J. F. F. Mendes, Adv. Phys. {\bf 51},
1079(2002).
\bibitem{Reviews3} M. E. J. Newman, SIAM Review {\bf 45}, 167(2003).
\bibitem{Reviews4} X. -F. Wang, Int. J. Bifurcation \& Chaos {\bf 12}, 885(2002).

\bibitem{Internet1} M. Faloutsos, P. Faloutsos and C. Faloutsos, Computer Communications Review {\bf 29},
251(1999).
\bibitem{Internet2} R. Pastor-Satorras, A. V\'{a}zquez and A. Vespignani, Phys. Rev.
Lett. {\bf 87}, 258701(2001).
\bibitem{Internet3} G. Caldarell, R. Marchetti and L. Pietronero, Europhys. Lett. {\bf 52}, 386(2000).

\bibitem{WWW1} R. Albert, H. Jeong and A. -L. Barab\'{a}si, Nature {\bf 401},
130(1999).
\bibitem{WWW2} B. A. Huberman, {\it The Laws of the Web} (MIT Press, Cambridge, 2001).

\bibitem{Social1} J. Scott, {\it Social Network Analysis: A Handbook} (Sage Publications, London,
2000).
\bibitem{Social2} S. Wasserman and K. Faust, {\it Social Network Analysis}
(Cambridge University Press, Cambridge, 1994).
\bibitem{Social3} F. Liljeros, C. R. Edling, L. A. N. Amaral, H. E. Stanley and Y. \AA berg, Nature
{\bf 411}, 907(2001).

\bibitem{Metabolic1} H. Jeong, B. Tombor, R. Albert, Z. N. Oltvai
and A. -L. Barab\'{a}si, Nature {\bf 407}, 651(2000).
\bibitem{Metabolic2} J. Padani, Z. N. Oltvai, B. Tombor, A. -L. Barab\'{a}si and E. Szathmary,
Nature Genetics {\bf 29}, 54(2001).
\bibitem{Metabolic3} J. Stelling, S. Klamt, K. Bettenbrock, S. Schuster and E. D. Gilles, Nature {\bf 420},
190(2002).

\bibitem{Foodwebs1} S. L. Pimm, {\it Food Webs} (University of Chicago Press, Chicago,
2002).
\bibitem{Foodwebs2} J. Camacho, R. Guimer$\grave{a}$ and L. A. N. Amaral, Phys.
Rev. Lett. {\bf 88}, 228102(2002).
\bibitem{Foodwebs3} R. J. Williams, E. L. Berlow, J. A. Dunne, A. -L. Barab\'{a}si and N. D. Martinez, PNAS {\bf
99}, 12913(2002).
\bibitem{Foodwebs4} J. A. Dunne, R. J. Williams and N. D. Martinez, PNAS {\bf 99}, 12917(2002).

\bibitem{Others1} Y. He, X. Zhu and D. -R. He, Int. J. Mod. Phys. B {\bf 18},
2595(2004).
\bibitem{Others2} T. Xu, J. Chen, Y. He and D. -R. He, Int. J. Mod.
Phys. B {\bf 18}, 2599(2004).
\bibitem{Others3} T Zhou, B. -H. Wang, P. -M. Hui and K. P. Chan, arXiv: cond-mat/0405258.
\bibitem{Others4} P. Sen, S. Dasgupta, A. Chatterjee, P. A. Sreeram, G. Mukherjee and S. S.
Manna, Phys. Rev. E {\bf 67}, 036106(2003).
\bibitem{Others5} J. R. Banavar, A. Maritan, and A. Rinaldo, Nature {\bf 399},
130(1999).
\bibitem{Others6} G. B. West, J. H. Brown and B. J. Enquist, Science {\bf 276}, 122(1997).
\bibitem{Others7} G. B. West, J. H. Brown and B. J. Enquist, Nature {\bf 400}, 664(1999).

\bibitem{BA} A. -L. Barab\'{a}si and R. Albert, Science {\bf 286},
509(1999).

\bibitem{Newman2001PRE} M. E. J. Newman, Phys. Rev. E {\bf 64},
016131(2001).

\bibitem{Newman2001PNAS} M. E. J. Newman, PNAS {\bf 98},
404(2001).

\bibitem{Fan2004} Y. Fan, M. Li, J. Chen, L. Gao, Z. Di and J. Wu, Int. J. Mod.
Phys. B {\bf 18}, 2505(2004);

\bibitem{Li2004}M. Li, Y. Fan, J. Chen, L. Gao, Z. Di and J. Wu, arXiv:
cond-mat/0409272.


\bibitem{MSN} G. Cs\'{a}nyi and B. Szendr\H{o}i, Phys. Rev. E
{\bf 69}, 036131(2004).

\bibitem{Directorships1} G. F. Davis and H. R. Greve, Am. J.
Sociol. {\bf 103}, 1(1997).

\bibitem{Directorships2} M. E. J. Newman, D. J. Watts and S. H. Strogatz, PNAS {\bf 99},
2566(2002).

\bibitem{WS1998} D. J. Watts and S. H. Strogatz, Nature {\bf
393}, 440(1998).


\bibitem{Women} A. Davis, B. B. Gardner and M. R. Gardner, {\it
Deep south} (University of Chicago Press, Chicage, 1941).

\bibitem{Myers2003} C. R. Myers, Phys. Rev. E {\bf 68},
046114(2003).

\bibitem{RDP} J. J. Ramasco, S. N. Dorogovtsev and R.
Pastor-Satorras, Phys. Rev. E {\bf 70}, 036106(2004).

\bibitem{Li2005} M. Li, J. Wu, D. Wang, T. Zhou, Z. Di and Y. Fan,
arXiv: cond-mat/0501655.

\bibitem{RAN1} T. Zhou, G. Yan, P. -L. Zhou, Z. -Q. Fu and B. -H.
Wang, arXiv: cond-mat/0409414.

\bibitem{RAN2} T. Zhou, G. Yan and B. -H. Wang, Phys. Rev. E {\bf
71}, 046141(2005).

\bibitem{SED} J. Laherr\`{e}re and D. Sornette, Eur. Phys. J. B {\bf
2}, 525(1998).

\bibitem{Ultrasmall} A. F. Rozenfeld, R. Cohen, D. ben-Avraham and
S. Havlin, Phys. Rev. Lett. {\bf 89}, 218701(2002).

\bibitem{Klemm} K. Klemm and V. M. Egu\'{i}luz, Phys. Rev. E
{\bf 65}, 036123(2002).


\bibitem{GT1} B. Bollob\'{a}s, {\it Modern Graph Theory} (Springer-Verlag
Publishers, New York, 1998).

\bibitem{GT2} J. -M. Xu, {\it Theory and Application of Graphs} (Kluwer
Academic Publishers, 2003).

\bibitem{Ameral2000} L. A. N. Amaral, A. Scala, M. Barth\'{e}l\'{e}my and H. E.
Stanley, PNAS {\bf 97}, 11149(2000).

\bibitem{Strogatz2001} S. H. Strogatz, Nature {\bf 410},
268(2001).

\bibitem{Bernard1988} H. R. Bernard, P. D. Killworth, M. J. Evans,
C. McCarty and G. A. Shelley, Ethnology {\bf 27}, 155(1988).

\bibitem{Lehmann2003} S. Lehmann, B. Lautrup and A. D. Jackson,
Phys. Rev. E {\bf 68}, 026113(2003).

\bibitem{Scala2000} A. Scala, L. A. N. Amaral and M.
Barth\'{e}l\'{e}my, arXiv: cond-mat/0004380.

\bibitem{Zhang2005} P. -P. Zhang, Y. He, T. Zhou, B. -B. Su, H. Chang, Y. -P. Zhou, B. -H. Wang and D. -R. He, Preprint.

\bibitem{Zhang2004} Y. Zhang, Y. He and D. -R. He, Bulletin of
APS, {\bf 49}(1), 1007(2004).

\bibitem{Davis1996} G. F. Davis, Corp. Govern. {\bf 4}, 154(1996).

\bibitem{Configuration} M. E. J. Newman, S. H. Strogatz and D. J.
Watts, Phys. Rev. E {\bf 64}, 026118(2001).

\bibitem{Manderbrot1983} B. B. Mandelbrot, {\it The fractal geometry of
nature} (Freeman, New York, 1983).

\bibitem{Mantegna1995} R. Mantegna, H. E. Stanley, Nature {\bf 376}, 46(1995).

\bibitem{Xie2004} Y. -B. Xie, B. -H. Wang, B. Hu and T. Zhou,
Phys. Rev. E {\bf 71}, 046135(2005).

\bibitem{Bak1994} P. Bak, {\it How nature works: the science of self-organized
criticality} (Freeman, New York, 1994).

\bibitem{Wang2001} B. -H. Wang and P. -M. Hui, Eur. Phys. J. B {\bf 20},
573(2001).

\bibitem{Newman2004} M. E. J. Newman, arXiv: cond-mat/0412004.

\bibitem{ex} Here, $m$-complete network means the networks of $m$
nodes fully connected to each other, which is denoted by $K_m$ in
mathematical literatures.

\bibitem{RateEquation} P. L. Krapivsky, S. Redner and F. Leyvraz, Phys. Rev.
Lett. {\bf 85}, 4629(2000).

\bibitem{Ravasz2003} E. Ravasz and A. -L. Barab\'{a}si, Phys. Rev. E {\bf 67}, 026112(2003).


\end{thebibliography}
\end{document}